# The Modeler-Schema Theory of Consciousness: Diffuse Awareness, Focal Experience, and a Saccadic Consistency Experiment

Frank Heile, Ph.D.

## Abstract

We propose that the Modeler-schema, a cybernetic control agent that monitors the brain's Modeler as it constructs and updates the World Model, is the minimal phenomenal subject. The theory distinguishes focal experience from diffuse awareness: focal attention selects focal-experience content, while diffuse awareness is the conscious presence of the full sensory field. The Human agent comprises three agents—Modeler, Controller, and Targeter—each paired with a schema agent; only the Modeler-schema generates qualia. It converts World Model contents into qualia for consistency checking and model refinement and, during saccades, may compare pre- and post-saccadic visual qualia across focal and non-focal regions and issue bottom-up attention requests for detected discrepancies. The proposed experiment tests the non-focal component of this comparison; a null result would leave focal-target checking and the broader theory viable. The theory addresses the Hard Problem through a physical identity unifying subject, process, and content within the Modeler-schema itself.

## Introduction

In this theory, an agent takes input, pursues internally represented goals, and interacts with the world or other agents; goals may be evolutionarily shaped or learned.

Four requirements motivate the proposed functional decomposition: modeling the body and environment, selecting actions, allocating scarce focal attention, and monitoring model-building accuracy. We assign these roles primarily to the Modeler, Controller, Targeter, and Modeler$_{schema}$, respectively. The **Human**[1] agent comprises the first three agents, each regulated by its own schema agent (defined in "Introduction to Cybernetic Regulation"):

- **Modeler**: builds and updates the World Model of the body and environment; supplies **Focal Target Information** to the Controller; and generates candidate bottom-up attention requests based on current sensory contents of the World Model, salience, and Controller-modulated goal priorities.

[1] Theory-specific terms, whether single words or multi-word phrases, appear in Title Case and **boldface** at their formal definition; earlier partial descriptions and later uses retain Title Case only.

- **Controller**: uses Focal Target Information to propose, evaluate, and initiate actions; issues task-driven attention requests based on task demands; and modulates goal priorities to bias what the Modeler treats as relevant.

- **Targeter**: because focal-attention capacity is limited, it integrates and prioritizes bottom-up attention requests from the Modeler and task-driven attention requests from the Controller, selecting the final focal attention targets that guide the Controller's actions.

The **World Model** has three interrelated components: the Concrete World Model built from current sensory inputs; the Memories of the Concrete World Model; and the Abstract World Model (Figure 1).

Our thesis is that the **Modeler$_{\textbf{schema}}$**—the agent that regulates the Modeler—is the minimal phenomenal subject—the smallest functionally identifiable agent in which conscious experience occurs; its quale-conversion and consistency-checking activities constitute experiencing, and its **Quale World Model** constitutes experienced content. The Modeler$_{schema}$ monitors World Model updates and performs qualia-based consistency checks, which may include pre- versus post-saccadic comparisons. When a discrepancy is detected, the Modeler$_{schema}$ prompts the Modeler to refine its processing and may also generate a bottom-up attention request to bring the discrepant content into focal attention for the Controller to examine.

We distinguish two possible forms of this cross-saccadic consistency check. Focal-target checking compares only qualia corresponding to focal targets, whereas full-field checking additionally compares non-focal diffuse-awareness qualia. The proposed experiment tests this additional full-field comparison; a sufficiently sensitive null result would reject it while leaving focal-target checking and the broader theory viable.

Cybernetic regulation alone is insufficient for consciousness; the Modeler$_{schema}$ additionally generates qualia and checks World Model coherence for refinement. These activities constitute the experiencing processes in the theory's identification of conscious experience with the Modeler$_{schema}$'s unity of subject, process, and content.

Two considerations motivate distinct but coupled World Model and Quale World Model.[2] First, task-usable information can remain available when corresponding experience is absent or degraded, as in aphantasia. More fundamentally, the World Model supports prediction and action, whereas the Quale World Model supports consistency checking and refinement using comparison-optimized qualia. The Modeler$_{schema}$ constructs the Quale World Model from Concrete World Model contents and Focal Target Information through Concrete, Memories of the Concrete, and Abstract quale-conversion processes. Its qualia-based processing influences behavior indirectly through model refinement or, under full-field checking, possible bottom-up attention requests.

We also propose fast-Modelers and fast-Controllers as evolutionary shortcuts enabling rapid actions that precede awareness. Before developing the theory's details, we clarify its level of analysis and cybernetic framework.

[2] Both considerations are developed in "Distinct Qualia from Distinct World Model Components," including the aphantasia evidence and the three quale-conversion processes.

At the cognitive-science level, this functional theory treats Modeler, Controller, Targeter, World Model, and schema agents as roles and information flows, not fixed anatomy. Neural implementations may be distributed and overlapping, provided these functional distinctions are preserved; specifying neural mechanisms is outside the present scope.

### Introduction to Cybernetic Regulation

Cybernetics supplies the idea that regulation requires modeling: Conant and Ashby's (1970) Good Regulator Theorem and Friston's (2010) Free Energy Principle formalize that effective regulators model what they regulate. Thus, if the Modeler constructs an explicit World Model, the $\text{Modeler}_{\text{schema}}$ must model that process.

The Human, Controller, Targeter, and Modeler are paired with the $\textbf{Human}_{\textbf{schema}}$, $\textbf{Controller}_{\textbf{schema}}$, $\textbf{Targeter}_{\textbf{schema}}$, and the previously defined $\text{Modeler}_{\text{schema}}$. Friston's predictive-processing framework maps mainly onto the regulated Modeler–World Model–Controller system, while precision weighting provides a partial analogue of the Targeter's attentional prioritization: the Modeler generates predictions, updates the World Model, and supplies relevant current and predicted information to the Controller through Focal Target Information; the Controller uses that information to guide action.

The $\text{Modeler}_{\text{schema}}$ regulates the Modeler in two ways: (a) controlling operating state and processing transitions and (b) using qualia to check World Model consistency. The Free Energy Principle can explain prediction-error minimization and behavior, but not the character of subjective experience. This theory adds a distinct mechanism: the $\text{Modeler}_{\text{schema}}$ generates qualia for consistency checks and uses discrepancies to refine Modeler processing.

## The Three-Part Structure of the World Model

In this theory, the World Model is the structured set of representations the Modeler constructs and maintains to supply Focal Target Information for Controller planning and action. It has three interrelated components:

- **Concrete World Model**: the dynamic, short-term representation updated by current sensory inputs, including recognized objects, bodily states, and emotions treated as internal sensory states. Its sensory contents are continuously available to diffuse awareness. When an object or event is no longer sensed, it is retained, at least temporarily, in the Memories of the Concrete World Model, preserving concrete-format continuity between current and remembered content.

- **Memories of the Concrete World Model**: short-, medium-, and long-term concrete-format records preserving past Concrete World Model content and storing imagined concrete objects, at least temporarily. Recalled memories can recreate Focal Target Information approximating the original target.

- **Abstract World Model**: short-, medium-, and long-term concepts and relations, from words tied to concrete objects to abstract ideas. It stores symbolic and conceptual regularities abstracted from the Concrete World Model and its Memories.

Figure 1 summarizes the Human World Model:

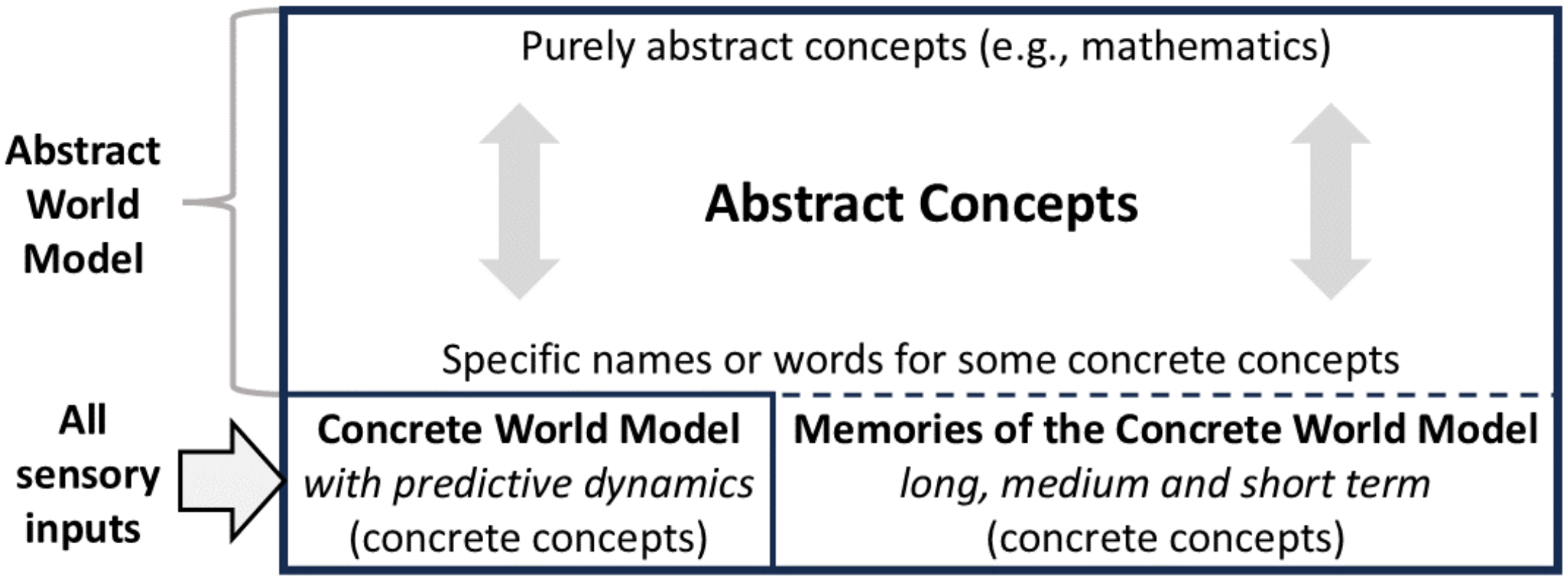

*Figure 1—The Human World Model*

Predictive-coding frameworks often treat memory as learned parameters or internal representations; this theory instead distinguishes actively updated Concrete, stored concrete-format, and abstract symbolic/conceptual representations, forming a continuum from concrete to abstract.

The World Model also encodes implicit physical and social regularities that let the Modeler predict object dynamics, other agents' behavior, and consequences of proposed actions or utterances for Controller planning.

Key terminology:

- **Focal attention** selects focal-experience targets.
- **Targets** are always focal attention targets: current, remembered, or imagined concrete objects, or abstract concepts.
- **Sensory stream** means a continuous flow of sensory data for one or more concrete targets or, more broadly, the full set of sensory streams supplying the Concrete World Model.
- **Abstract stream** means a discrete, finite, ordered sequence of abstract targets, such as the concepts represented by words in a sentence.
- **Diffuse attention** corresponds to diffuse awareness, discussed in "Diffuse Attention and Diffuse Awareness."

## The Three-Agent Model

Figure 2 illustrates the Human's three agents—Modeler, Controller, and Targeter—and their connections.

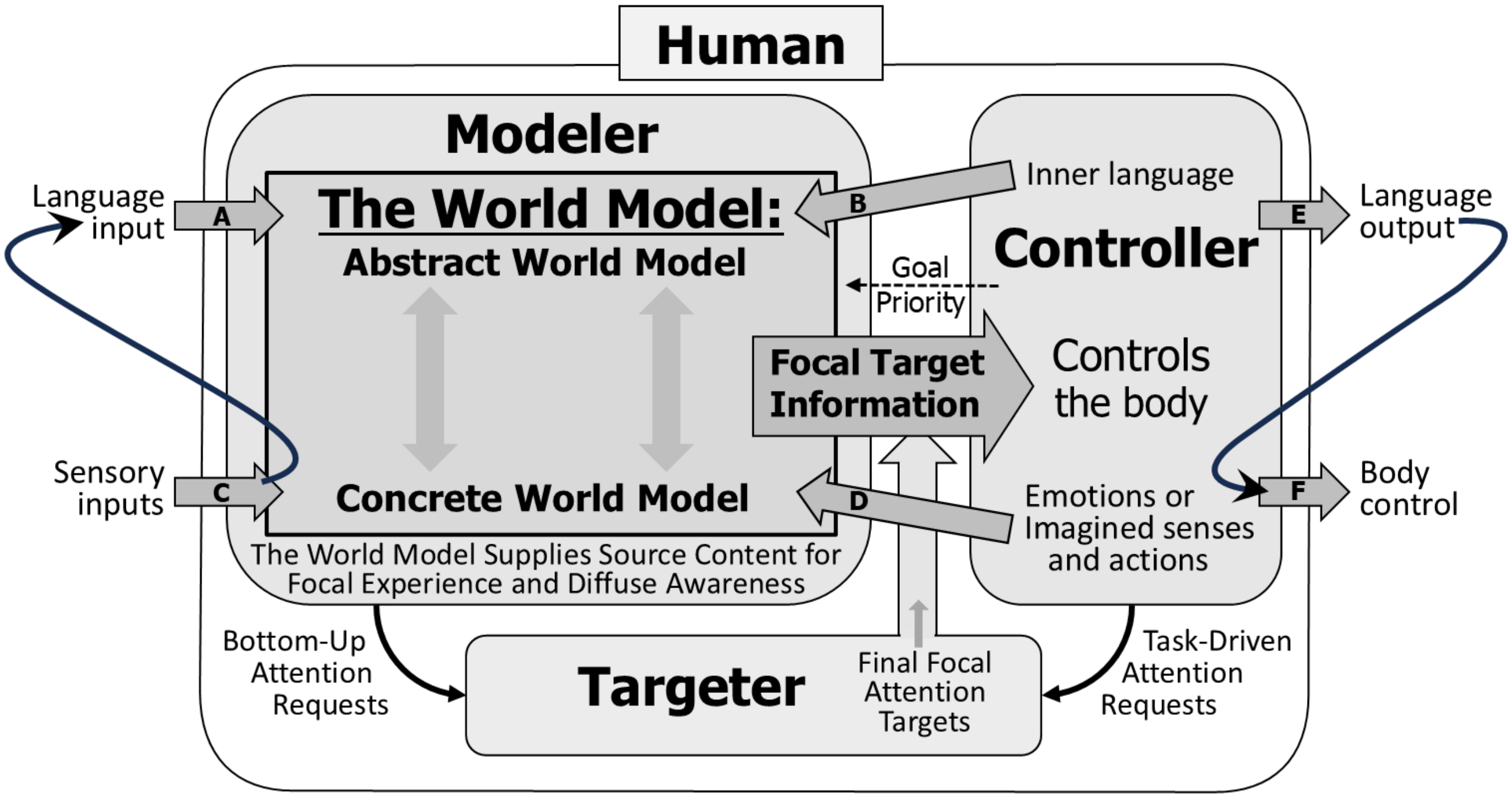

*Figure 2—The Three-Agent Model of the Human*

The Modeler receives inputs through arrows A–D, uses them to construct and update the World Model, and sends Focal Target Information to the Controller. The dashed arrow carries Controller goal priority signals that bias the Modeler's bottom-up attention requests. The Controller acts through arrows B, D, E, and F but can update the World Model only through the Modeler. The Targeter prioritizes bottom-up attention requests from the Modeler and task-driven attention requests from the Controller; its selections become final focal attention targets and determine the Focal Target Information stream's content.

Human sensory inputs enter the Modeler through arrow C; motor outputs leave the Controller through arrow F. Arrows A, B, and E respectively carry language input, inner language, and language output as abstract content in the brain's internal format, not any natural language. Inner language on B may be a monologue, such as rehearsed speech, or a dialogue: the Controller poses a question, and the Modeler returns a retrieved, inferred, or predicted answer through Focal Target Information. The Modeler processes language content on A and B to update the Abstract World Model; the Controller constructs language output on E. That output may be monitored internally through B or externally when speech or writing produced through arrows E and then F re-enters through C and then A. Arrow D carries Controller-generated content—emotions, imagined sensory objects, or imagined actions. Emotions update the Concrete World Model as internal sensory states; imagined sensory objects update the Memories of the Concrete World Model; and imagined actions prompt the Modeler to predict likely consequences and return those predictions to the Controller through Focal Target Information.

The Controller can report the content of focal targets using information from the Focal Target Information stream. In contrast, diffuse awareness is conscious but not directly reportable: the Controller can describe peripheral visual content only after it becomes a focal target.

## Whole Agents and Consciousness Attribution

A **Whole Agent** comprises a functional agent, its schema agent, and any symbolic World Model representations of that agent. The **Controller$_{\textbf{whole}}$** comprises the Controller, Controller$_{schema}$, and two symbolic representations:

- **Body$_{\textbf{model}}$**: encodes bodily state and dynamics in the Concrete World Model and its Memories, and may incorporate frequently used tools.
- **I/Me/My$_{\textbf{model}}$**: represents abstract identity, personal history, and conceptual self in the Abstract World Model.

Together, the Body$_{model}$ and I/Me/My$_{model}$ represent the Controller$_{whole}$ within the World Model. These stand-ins enable the Controller$_{whole}$ to plan actions and communicate about itself.

The **Modeler$_{\textbf{whole}}$** and **Targeter$_{\textbf{whole}}$** consist only of their primary and corresponding schema agents, with no symbolic representations. The **Human$_{\textbf{whole}}$** contains the Controller$_{whole}$, Modeler$_{whole}$, and Targeter$_{whole}$; therefore, it uniquely combines the Controller$_{whole}$'s symbolic representations with the Modeler$_{schema}$ functionally incorporated through the Modeler$_{whole}$.

Figure 2's primary-agent interactions also apply to the corresponding Whole Agents because schema agents can generate or influence primary-agent outputs and share some capabilities; each pair is tightly coupled.

Figure 3 depicts the Human, Human$_{whole}$, and Human$_{schema}$ in panels (a), (b), and (c), respectively. The red-outlined rectangles in panels (b) and (c) mark the Modeler$_{schema}$. This nested architecture supports the claim: conscious experience occurs only in the Modeler$_{schema}$ and is attributed only to agents that functionally include it.

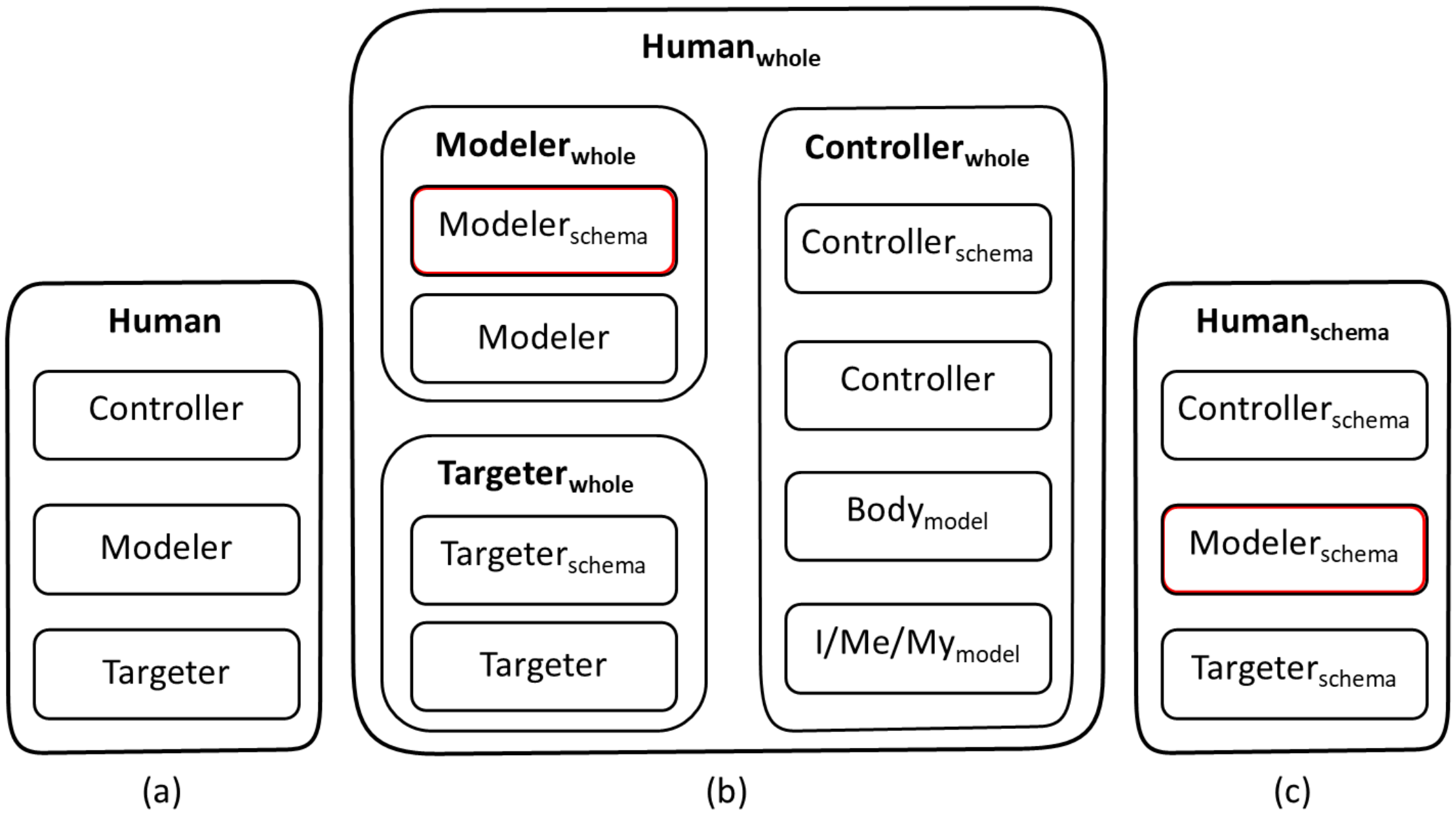

*Figure 3—Composition of the Agents and Whole Agents*

The Modeler$_{schema}$ is the minimal phenomenal subject: all conscious experience occurs within it. The Modeler$_{whole}$, Human$_{schema}$, and Human$_{whole}$ functionally incorporate the Modeler$_{schema}$ and may therefore be described derivatively as conscious; they are not independent subjects of experience. All other agents are nonconscious.

Because the Controller$_{whole}$ owns the body, acts, and uses language, it identifies itself with the Human, which it takes to be conscious. The theory instead identifies the Modeler$_{schema}$—not the Controller$_{whole}$, Body$_{model}$, or I/Me/My$_{model}$—as the conscious subject.

## Diffuse Attention and Diffuse Awareness

We distinguish "awareness" from "experience" by the type of attention involved:

- **Awareness** corresponds to diffuse attention: broad, unfocused, and all-encompassing across the multisensory field.
- **Experience** corresponds to focal attention: directed at one or a few objects or concepts.[3]

Visual diffuse awareness is field-wide but not spatially uniform: resolution, precision, and feature detail vary across the sensory field and are generally enhanced for focal targets—yet peripheral objects can still appear stable and well formed despite being represented with reduced detail.

[3] We sometimes use "experience" inclusively to encompass both focal experience and diffuse awareness.

William James (1890, p. 402) wrote [italics in original]:

> "Millions of items of the outward order are present to my senses which never properly enter into my experience. Why? Because they have no *interest* for me. *My experience is what I agree to attend to*."

We interpret James's "millions of items… present to my senses" as diffuse awareness[4] and "*My experience is what I agree to attend to*" as focal experience. In our terms, "no *interest* for me" corresponds to the Modeler's *assessment* process, introduced below: content *assessed* as "*not interesting*" remains in diffuse awareness, whereas "*interesting*" content may generate a bottom-up attention request and may result in a focal experience.

Treisman and Gelade (1980) made a similar observation:

> "When we open our eyes on a familiar scene, we form an immediate impression of recognizable objects, organized coherently in a spatial framework."

Their "immediate impression" is the conscious outcome of what they later call preattentive feature registration; in this theory, that corresponds to Modeler processing that constructs the Concrete World Model supporting diffuse awareness.

To compare visual diffuse awareness with visual focal experience, try this exercise:

- Direct your central vision to an object and keep focusing on it.
- Without moving your eyes, take in the entire forward visual field, out to its edges.
  - Attend to the whole field without fixating on any item.
  - This is diffuse awareness.
- While still focused on the same object, try identifying a peripheral object without moving your eyes, noting its color or shape.
  - Notice the greater effort compared with the earlier holistic diffuse awareness.

This exercise should reveal two things.

- Selecting or describing a peripheral object requires peripheral visual focal attention,[5] unlike relatively effortless diffuse awareness.[6]
- The $\text{Controller}_{\text{whole}}$ can report peripheral objects only when selected as focal targets.

---

[4] James includes all senses here, not just vision.
[5] Peripheral visual focal attention means focusing on peripheral objects without eye movement and corresponds to standard covert attention.
[6] The main challenge is to resist directing peripheral visual focal attention to specific items within diffuse awareness.

Thus, the $\text{Controller}_{\text{whole}}$ accesses only Concrete World Model content selected as focal targets, while diffuse-awareness content remains conscious but not directly reportable. Accordingly, diffuse awareness is best located within the $\text{Modeler}_{\text{whole}}$ rather than the $\text{Controller}_{\text{whole}}$. For now, we posit that both diffuse awareness and focal experience occur within the $\text{Modeler}_{\text{schema}}$.

## Functions of the Modeler: *Understanding*, *Assessing*, and *Updating*

The Modeler performs three overlapping functions: *understanding*, *assessing*, and *updating*.[7]

*Understanding* involves recognizing patterns in sensory and abstract inputs and integrating them into the World Model. This includes interpreting speech and text[8] and building a unified, predictive model of physical and social environments.

*Assessment* evaluates the world state relative to Controller goal priorities and can mark events as "*interesting*," potentially generating bottom-up attention requests. *Assessment* handles single features or disjunctions such as "red or vertical," but may treat conjunctions such as "red and vertical" as disjunctions, prompting sequential inspection. Treisman and Gelade (1980) attributed such conjunction errors to faulty peripheral object construction (i.e., *understanding*); we locate them in *assessment*.

The Concrete World Model is continuously *updated* from current sensory input while we are awake; the Focal Target Information stream is *updated* as needed to keep the Controller informed about its tracked targets. All *updated* World Model content is stored in short-, medium-, or long-term memory as concrete or abstract representations.

These processes may also generate Appraisal Emotions, discussed next.

### Appraisal Emotions

Ordinary emotions, such as *fear* or *anger*, generated by the Controller can directly affect its action selection and also enter the Concrete World Model, where they are represented as internal sensory states. We now propose a second class, Appraisal Emotions, generated within the $\text{Modeler}_{\text{whole}}$, that can also affect behavior through the $\text{Controller}_{\text{whole}}$.

**Appraisal Emotions**[9] are affective states, typically expressed adjectivally or adverbially, generated by Modeler functions—*understanding*, *assessing*, and *updating*—or by $\text{Modeler}_{\text{schema}}$ processing, especially quale generation. They track how the $\text{Modeler}_{\text{whole}}$ processes a target or event. Appraisal Emotions affect the $\text{Controller}_{\text{whole}}$ when attached to focal targets, but are experienced only as Appraisal Qualia.

---

[7] Formatting note: Modeler-function verbs and their inflected forms appear in italics (e.g., *understand*, *understanding*, *understood*). Emotion nouns are also italicized (e.g., *anger*, *fear*).
[8] Speech and text input is represented by the curved black arrow from arrow C (sensory inputs) to arrow A (language input) in Figure 2.
[9] Drawing on Arnold's (1960) account of emotion as arising from appraisals of one's environment. Here, Appraisal Emotions arise as inward-directed appraisals of $\text{Modeler}_{\text{whole}}$ processing.

Examples of Modeler-generated Appraisal Emotions include "*understood*," "*not understood*," "*confused*," "*suddenly*," and "*fluently*" from *understanding*, and "*interesting*" or "*not interesting*" from *assessing*.[10] The Modeler$_{schema}$ may generate similar appraisals, plus qualia-related appraisals such as "*vivid*," "*striking*," "*faint*," or "*absent*."

The "*understood*" versus "*not understood*" contrast is easy to experience. Compare:

- **Sentence 1**: The hundred always one sum in the any is degrees eighty triangle angles of.
- **Sentence 2**: The sum of the angles in any triangle is always one hundred eighty degrees.

Sentence 1 is likely experienced as "*not understood*" and Sentence 2 as "*understood*." (Sentence 1 scrambles Sentence 2's words.) Beyond Sentence 1's longer reading time, that felt difference is an Appraisal Quale. These "*understood*" and "*not understood*" labels are English interpretations of associated nonlinguistic Appraisal Emotion signals, which the Controller$_{whole}$ can learn to associate with corresponding concepts and words used in reports.

## The Four Types of Qualia

Qualia are internal representations constituting the content of experience within the Modeler$_{schema}$ and available only to it. They are model contents used for consistency checking, not a non-physical substance. The theory distinguishes four types:[11]

- **Sensory Qualia**: perceptual content and emotions treated as internal senses, drawn from the Concrete World Model. They constitute the content of focal sensory experience and diffuse awareness of the multisensory field.
- **Memory-Based Sensory Qualia**: sensory content from remembered or imagined objects, such as a recalled face or an imagined new object,[12] drawn from Memories of the Concrete World Model.
- **Cognitive Qualia**: abstract or symbolic content drawn from the Abstract World Model, including inner language and structured conceptual thought.
- **Appraisal Qualia**: Appraisal Emotion content reflecting how a target or event is processed, including "*not understood*," "*vivid*," "*absent*," "*suddenly*," and "*fluently*."

Together, these four types of qualia constitute the Quale World Model.

---

[10] Formatting note: Appraisal Emotions are italicized and quoted (e.g., "*not understood*," "*suddenly*").
[11] Some authors restrict "qualia" to sensory qualities and sensory imagery; here, the term encompasses all four types of experienced content.
[12] Imagined objects are stored there at least temporarily because they become focal targets and may later be recalled.

Only the Modeler$_{\text{schema}}$ contains these qualia and performs the processes that constitute conscious experiencing. The Controller$_{\text{whole}}$ receives only Focal Target Information and associated Appraisal Emotion signals, not qualia.

Metacognition and introspective awareness are Controller$_{\text{whole}}$ narratives from Focal Target Information and Appraisal Emotion signals, not separate quale types; the corresponding focal-target qualia occur only within the Modeler$_{\text{schema}}$.

## Distinct Qualia from Distinct World Model Components

All experienced content consists of qualia in the Quale World Model, constructed and updated by the Modeler$_{\text{schema}}$. Diffuse-awareness qualia derive from the Concrete World Model, while focal-target qualia derive from whichever World Model component supplies the current targets. This asymmetry reflects current functional relevance. The Concrete World Model represents the presently sensed body and environment, and any part of its sensory field can become behaviorally relevant when the Modeler generates a bottom-up attention request in response to sensory change, salience, or changed Controller goal priorities; its full sensory field is accordingly available for diffuse-awareness quale conversion. For the Memories of the Concrete and Abstract World Models, only focal-target content is immediately relevant to the Controller$_{\text{whole}}$; when such content is selected, it enters Focal Target Information and undergoes the corresponding quale conversion. Thus, the Quale World Model represents the full current sensory field together with all currently focal content from the Memories of the Concrete or Abstract World Models. We therefore propose three quale-conversion processes: Concrete, Memories of the Concrete, and Abstract.

Appraisal Qualia do not need a fourth quale-conversion process because they are qualities associated with content from any World Model component. Their Appraisal Emotion signals are converted by that content's quale-conversion process and are indirectly reportable only when attached to focal targets received by the Controller$_{\text{whole}}$.

The Concrete World Model is continuously *updated* by sensory input and always available to the Modeler$_{\text{schema}}$. Concrete visual diffuse-awareness and focal-target qualia are vivid, stable, and relatively consistent across individuals, presumably because accurate perceptual consistency checking is under strong evolutionary pressure.[13] In contrast, the Memories of the Concrete World Model and Abstract World Model are available only as focal targets, and their qualia reportedly vary much more across individuals.

Pearson (2019) documents visual recall and imagery ranging from absent to highly vivid. Jacobs, Schwarzkopf, and Silvanto (2017) found an aphantasic participant with near-normal visual working-memory performance despite no reported visual imagery; she reported "simply 'knowing' what the correct answer is." We propose that recalled visual information remained available to the Controller$_{\text{whole}}$ through Focal Target Information despite an absent or severely degraded visual quale.

[13] A rare, easily describable exception is grapheme–color synesthesia, in which a letter or number evokes an additional color quale from its abstract identity rather than its physical color.

Abstract qualia also vary: Alderson-Day and Pearson (2023) document verbal and nonverbal inner speech, and Gregory (2026) discusses signed inner speech with visual or primarily kinesthetic phenomenology.

Together, these observations motivate three distinct functional pathways to experienced content—from (a) current sensory content, (b) remembered or imagined sensory content, and (c) abstract or symbolic content—which we implement as three corresponding quale-conversion processes:

| Component | Qualia Produced by the Corresponding Quale-Conversion Process |
|---|---|
| Concrete World Model | Produces stable, vivid sensory qualia under strong evolutionary pressure because the Modeler$_{schema}$ relies on them for Concrete World Model refinement and consistency checks. |
| Memories of the Concrete World Model | Produces recalled or imagined sensory qualia ranging from highly vivid to absent, reflecting individual differences in qualia generation for sensory recall and imagination. |
| Abstract World Model | Produces qualia for symbolic or conceptual content, including inner language and abstract reasoning, with high variability across individuals and language-acquisition modalities. |

*Table 1—The Quale-Conversion Processes for the Three World Model Components*

Thus, the greater variability of Memory-Based Sensory and Cognitive Qualia may reflect weaker evolutionary constraints on their phenomenal character.

## Visual Dynamics of the Concrete World Model

Controller$_{whole}$ action requires a stable World Model. Vision poses a special problem: saccades occur several times per second and last only tens of milliseconds, yet the visual world appears stable and continuous.

Burr and Morrone (2012) describe two complementary visual systems:

- **Transient Saccadic Process**:[14] a fast visual mechanism in eye-centered, retinotopic[15] coordinates that preserves continuity across saccades and supports rapid visual judgments (≈200 ms).
- **Stable spatial map**: a slower 3D multisensory, spatiotopic[16] body/world-centered representation constructed in dorsal pathways (≈500 ms).

Together, these systems support stable perception. During a saccade, the Transient Saccadic Process holds focal pre-saccadic output, helping suppress experience of eye movement, then aligns it with post-saccadic focal input. The stabilized output enters the stable spatial map, updating the focal Concrete

[14] "Transient Saccadic Process" is our term for what the article calls "transient systems."
[15] Retinotopic maps (e.g., V1–V4) are retina-centered.
[16] Spatiotopic maps are world-centered.

World Model smoothly.[17] Between saccades, focal visual input passes through the Transient Saccadic Process, while non-focal input enters the stable spatial map directly.

We treat Burr and Morrone's stable spatial map as functionally equivalent to our multisensory Concrete World Model, which integrates spatial and object-level information for sensory-guided action. The fast-Modelers and fast-Controllers mentioned earlier could explain how some actions precede awareness: a fast visual Modeler can remain retinotopic through the Transient Saccadic Process and directly drive a fast retinotopic Controller to act in ≈200 ms without requiring full Concrete World Model construction, whereas the full Modeler–Controller$_{whole}$ pathway can produce action in ≈500 ms. Analogous fast systems may support time-critical auditory processing and trained performance in sports or music.

## Optimizing Perception and Visual Consistency Checks

The Modeler$_{schema}$ monitors World Model accuracy by (a) generating qualia, (b) checking their consistency, and (c) using discrepancies to optimize Modeler construction. It receives the Concrete World Model's full current sensory contents and Focal Target Information, converting them into comparison-optimized diffuse-awareness and focal-target qualia.

A saccade typically shifts an attended object from peripheral to foveal vision, enabling comparison of pre- and post-saccadic qualia to test the Modeler's compensation for representational differences, such as peripheral cone density being about 5% of central density (Wells-Gray et al., 2016). The Transient Saccadic Process retains, aligns, and compares focal visual output, which remains focal data rather than comparison-optimized qualia. The Modeler$_{schema}$ may then compare the corresponding object-view qualia to refine Modeler processing.

When such a cross-saccadic comparison detects a difference between an object's pre- and post-saccadic qualia, the Modeler$_{schema}$ may also generate the Appraisal Emotion "*surprised*," thereby flagging possible inaccuracies in the Transient Saccadic Process. Such discrepancies may help calibrate peripheral-to-foveal compensation in infancy; later, they may signal modeling errors, new vision problems, or genuine object changes.

## The Sensory Data Pipeline

The following pipeline specifies how the Transient Saccadic Process, Concrete World Model, and Quale World Model interact.

Only focal visual targets pass through the Transient Saccadic Process, while the full sensory field contributes to the Concrete World Model. Thus, within the Modeler, the Transient Saccadic Process compares only focal visual targets across saccades, whereas the Modeler$_{schema}$ may additionally compare diffuse-awareness qualia across the full visual field.

[17] Melcher and Colby (2008), discussed later in "Differentiating Step 3c vs. full-field Step 5c Bottom-Up Attention Requests," support the claim that only focal visual concrete information is remapped.

The experiment tests this pipeline. Steps 1–4 are Modeler processing, and Step 5 is $\text{Modeler}_{\text{schema}}$ processing. The fixation target is foveal focal; the saccade target is initially peripheral focal and becomes foveal afterward. Steps 1 and 2 are concurrent.

Step 1: Visual *understanding* and *assessment*.

Data from the full visual sensory stream are initially *understood* and *assessed* in retinotopic coordinates. Focal visual data are routed to the Transient Saccadic Process in Step 3; the full visual stream continues toward construction and *updating* of the visual Concrete World Model in Step 4.

Step 2: Non-visual sensory *understanding* and *assessment*.

Non-visual sensory data are *understood* and *assessed* for construction and *updating* of the Concrete World Model in Step 4, where non-visual focal targets can also be integrated with corresponding focal visual targets.

Step 3: Transient Saccadic Process for focal visual targets.

The Transient Saccadic Process uses retinotopic coordinates and processes only focal visual targets.

Step 3a: Before the saccade.

Focal visual input flows continuously through the Transient Saccadic Process, refreshing its output.

Step 3b: During the saccade.

The Transient Saccadic Process holds its focal visual output at pre-saccadic values.

Step 3c: After the saccade.

The Transient Saccadic Process aligns post-saccadic focal input at the landing region—especially the now-foveal saccade target—with that target's pre-saccadic peripheral focal data before normal processing resumes. This alignment preserves target continuity and can extend to other focal targets. Step 3c may also produce a saccade-vector estimate for Step 5c.

Step 4: *Updating* the Concrete World Model and the Focal Target Information stream.

All sensory data are *understood* and *assessed* in Steps 1 and 2 before *updating* the Concrete World Model in Step 4. Focal visual targets additionally pass through Step 3; its output is integrated with relevant non-visual inputs into stable, world-based focal representations supplied to the Controller through Focal Target Information and used to *update* corresponding Concrete World Model portions.

Non-focal visual and all non-visual information from Steps 1 and 2, together with stabilized focal visual representations from Step 3, *update* the Concrete World Model's current sensory contents, including non-focal content for diffuse-awareness qualia. These contents and Focal Target Information are supplied to Step 5.

Step 5: Modeler$_{schema}$ quale conversion and visual cross-saccadic consistency checking.

Non-visual Concrete quale conversion continues to update non-visual diffuse-awareness qualia. For focal targets not derived from current visual Concrete World Model content, the appropriate quale-conversion process continues to construct focal-target qualia from Focal Target Information.

However, during saccades we propose that all current visual Concrete World Model content, whether focal or non-focal, follows the special update sequence described in Steps 5a–5c. This sequence updates the current-sensory visual portion of the Quale World Model, including, when appropriate, focal-target qualia corresponding to visual Focal Target Information.

Step 5a: Before the saccade.

The Concrete quale-conversion process continuously updates the current-sensory visual portion of the Quale World Model normally.

Step 5b: During the saccade.

The current-sensory visual portion of the Quale World Model is held at pre-saccadic values. This normally prevents intra-saccadic visual changes from directly updating current-sensory visual qualia; Step 3b likewise holds focal visual output fixed.

Step 5c: After the saccade.

The continuously *updated* visual Concrete World Model already contains visual content consistent with the new post-saccadic eye position. The Concrete quale-conversion process can begin generating candidate post-saccadic diffuse-awareness visual qualia, optimized for consistency comparison, from that content.[18]

Step 5c cannot replace the held pre-saccadic visual qualia all at once because each must remain available for comparison. Using Step 3c's saccade vector, the Modeler$_{schema}$ matches each candidate post-saccadic quale generated from the updated visual Concrete World Model with its corresponding pre-saccadic quale held in the Quale World Model. It compares each pair before replacing the held quale with the candidate.[19]

[18] As current-sensory visual qualia are updated, corresponding visual focal-target qualia are also updated when appropriate.

[19] This comparison is functionally non-local: each post-saccadic location is compared with its pre-saccadic location displaced by the saccade vector. This cognitive-level claim does not specify neural implementation; the correspondence problem may be solved more locally or efficiently. For example, Step 5c might use remapping like Step 3c's focal realignment rather than independently comparing displaced locations.

> If a candidate post-saccadic peripheral quale is inconsistent with its displaced pre-saccadic counterpart, the Modeler$_{schema}$ treats the discrepancy as a possible Modeler error or evidence of a peripheral non-focal change. It may then generate a bottom-up attention request and attach an Appraisal Emotion such as "*surprised*" to that content.

The remaining uncertainty is the scope of this Step 5 cross-saccadic quale comparison.

### Testing Whether Step 5 Uses Full-Field or Focal-Target Information

The architecture described above, which we call full-field Step 5, performs cross-saccadic consistency comparison across the full current-sensory visual Quale World Model, including non-focal diffuse-awareness qualia. We propose a simpler focal-target Step 5, which still holds and updates the full current-sensory visual Quale World Model but compares only visual focal-target qualia from Focal Target Information. The experiment tests whether Step 5 comparison extends beyond focal targets.

For focal-target Step 5, the Modeler's Step 3c has already identified and aligned each focal target across the saccade. The Modeler$_{schema}$ can therefore monitor the corresponding visual focal-target qualia for a sudden post-saccadic discrepancy without independently solving the cross-saccadic correspondence problem. Because these qualia are optimized for consistency comparison, focal-target Step 5c may detect a content discrepancy even when Step 3c has established that the pre- and post-saccadic inputs represent the same target. Across repeated saccades, this could compare peripheral and foveal representations of targets from many locations throughout the visual field, providing broad peripheral-to-foveal calibration over time.

Under both architectures, non-focal diffuse-awareness visual qualia are held during the saccade.[20] Focal-target Step 5 compares only visual focal-target qualia and then updates held non-focal qualia without comparison. Full-field Step 5c additionally compares held non-focal qualia with candidate post-saccadic qualia from the visual Concrete World Model, using Step 3c's saccade vector before updating them and allowing bottom-up attention requests for previously non-focal changes.

Outside of saccades, Modeler processing can detect goal-relevant or sufficiently salient changes, including changes whose transitions begin before a saccade or continue after it. Thus, focal-target Step 5 lacks only the ability to compare non-focal pre- and post-saccadic visual content when the entire transition of a permanent change occurs during the saccade.

Focal-target Step 5 is simpler because it relies on Step 3c's focal correspondences and requires no non-focal alignment or comparison. Detecting a non-focal change during a saccade after excluding Steps 1, 3c, and 4 would support full-field Step 5; a sufficiently sensitive null result would reject it while leaving focal-target Step 5 indistinguishable from Step 3c or $Controller_{whole}$ processing and the broader theory viable but largely untested by this experiment.

[20] The author's phenomenology is consistent with this proposal: when alternating saccades between two nearby focal targets and briefly pausing on each, the diffuse visual field appears stable while fixation seems to jump between them. By contrast, smoothly tracking a moving pencil tip between the same locations makes the peripheral field appear to slide back and forth.

## The Proposed Experiment

The experiment tests full-field Step 5 cross-saccadic consistency checking. Steps 1–4 are Modeler processing; Step 5 is $Modeler_{schema}$ processing. If a non-focal peripheral object changes during a saccade and the resulting bottom-up attention request cannot be attributed to Steps 1, 3c, or 4, full-field Step 5c is the only candidate source within the proposed architectures.

Participants may report attention capture without identifying the changed feature because an object becoming focal only post-saccade may lack a pre-saccadic focal representation for $Controller_{whole}$ comparison. They should be instructed to identify the object that seemed to change, even if only as a hunch, and report the changed feature when possible. Frequent feature identifications may therefore indicate retained pre-saccadic focal representations and should be minimized by varying features and positions across trials or shortening pre-saccadic viewing.

Step 1 detection is difficult because the image sweeps rapidly across the retina during a saccade. Accordingly, object changes are designed to occur slowly relative to instantaneous eye velocity. Because saccade velocity follows a bell-shaped curve, size and position changes use the same smooth temporal profile, scaled to a small fraction of instantaneous eye velocity; color changes use that profile in color space. Additional methods described below minimize Step 1-generated bottom-up attention requests and distinguish any that remain.

Step 3c aligns the intended saccade target with the actual landing point; when misalignment occurs, a corrective saccade may follow, or the Modeler's Transient Saccadic Process may remap and align the visual data to compensate (Burr & Morrone, 2012). Prior visual research and experimental controls help distinguish between Step 3c and full-field Step 5c contributions.

Step 4 is excluded below on both functional and representational grounds: it lacks pre-saccadic diffuse visual content, and a mechanism adding the required comparison would effectively be $Modeler_{schema}$ processing.

To distinguish these candidate sources, the experiment could use accurate eye trackers and a high-resolution, high-refresh-rate, high-bit-depth monitor configured as shown in Figure 4 (letter labels appear only during the response phase).

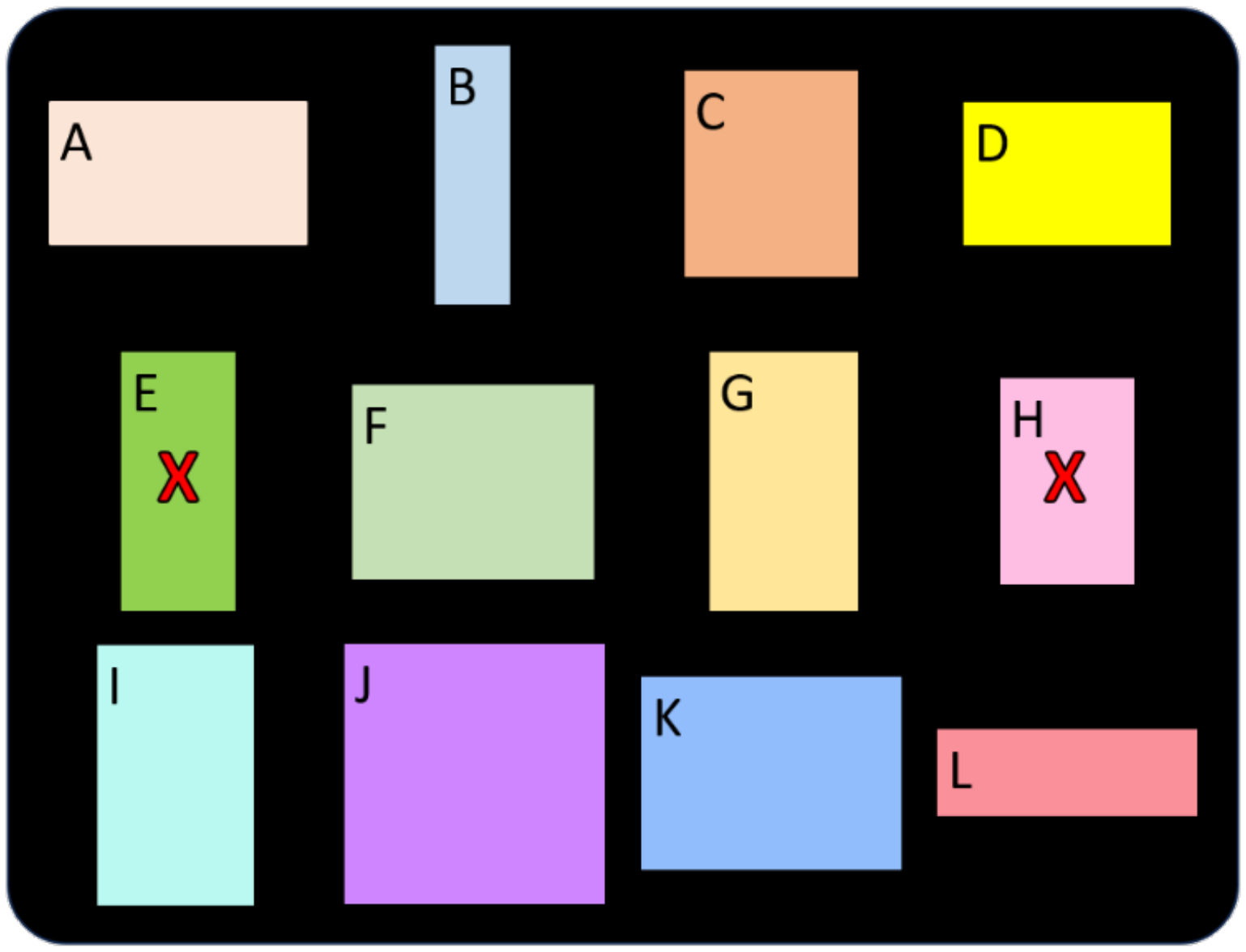

*Figure 4—Illustrative Experimental Display Layout*

The participant fixates on one red X and saccades to the other on cue. Before each saccade, the fixation X is foveal focal and the goal X peripheral focal, becoming foveal after the saccade. The changing object is never X-bearing, minimizing direct $\text{Controller}_{\text{whole}}$ comparison. If successful, changed peripheral objects become focal only after the saccade, supporting full-field Step 5.

During each saccade, zero, one, or more peripheral objects may subtly change in size, position, or color. Object size may increase or decrease horizontally or vertically while its center remains fixed. Position changes can be vertical or horizontal and may be coordinated with slight shifts in the destination target X. A change can be permanent, persisting after the saccade, or temporary, returning to baseline by saccade end.

Two background conditions are used: black, producing high-contrast edges, and mid-gray. Because Burr and Morrone (1996) found different temporal responses and levels of saccadic suppression for luminance-modulated and equiluminant chromatic stimuli, the experiment includes high-luminance-contrast and near-isoluminant display conditions. All element colors use a near-isoluminant palette and, on mid-gray, are also approximately luminance-matched to the background. These measures reduce luminance-based signals that could generate Step 1 bottom-up attention requests.

Full-field Step 5 should continue detecting permanent changes even when luminance contrast is minimized because they should still produce distinct qualia. The key pattern would be fewer temporary-change detections on mid-gray but preserved permanent-change detections. A follow-up could use per-observer peripheral isoluminance calibration to further suppress early luminance-based detections.

Color changes primarily alter object features; size and position changes also alter spatial relationships and object boundaries. Greater sensitivity to permanent than temporary changes in either class would support pre-/post-saccadic consistency comparison; finding the effect in both would strengthen the evidence.

As a control, no-saccade trials with identical timing provide a baseline for peripheral change reports and false alarms. Participants should identify temporary changes more often in these trials because the changing object can become a peripheral focal target while the change unfolds.

### Differentiating Step 1 vs. full-field Step 5c Bottom-Up Attention Requests

Temporary non-focal changes primarily test Step 1-generated bottom-up attention requests because they return to baseline before the saccade ends, leaving no pre-/post-saccadic discrepancy. For matched peak magnitude, temporary changes also have higher rates of change than permanent ones. Thus, Step 1 should favor temporary changes; full-field Step 5c should favor permanent changes. The strongest evidence for full-field Step 5 would be detection of permanent but not matched-magnitude temporary changes.

If temporary changes are detected, paired trials present a permanent change on one object and a temporary change of equal peak magnitude on another. Trials reporting both changes favor neither source. Full-field Step 5 is supported if the permanent change more often generates a bottom-up attention request, because Step 5c detects only permanent changes whereas Step 1 should favor the faster-changing temporary changes. Mid-gray should further reduce Step 1 detections for size and position changes while preserving full-field permanent-change detections.

Because high-magnitude temporary changes, such as brief flashes, can be consciously perceived despite saccadic suppression (Honda, 1989), each participant’s intrasaccadic detection threshold should be measured. The main experiment should use temporary changes well below that threshold, minimizing consciously visible intra-saccadic changes.

### Differentiating Step 3c vs. full-field Step 5c Bottom-Up Attention Requests

Melcher and Colby (2008) argue that predictive remapping is selective for salient, attended objects, not the unattended remainder of the scene. In our terms, Step 3c should therefore not remap diffuse-awareness content.

Research on intrasaccadic target shifts and nearby landmarks further defines this boundary. For example, small intrasaccadic target shifts are usually unreported even when corrective saccades recenter the fovea, suggesting that Step 3c generates no bottom-up attention request for them (Bridgeman, Hendry, & Stark, 1975). Nearby landmarks can also bias perceived stability (Deubel, Koch, & Bridgeman, 2010).

Together, these findings suggest that Step 3c processing is limited to focal-target regions and their immediate surroundings. This limit is critical: once Steps 1 and 4 are excluded, a bottom-up attention request caused by a small permanent change in a non-target peripheral object outside those regions cannot come from the $\text{Controller}_{\text{whole}}$ or Step 3c. Within the two proposed Step 5 architectures, it must come from full-field Step 5c, which performs cross-saccadic comparison of non-focal diffuse-awareness qualia.

To dissociate the mechanisms, we examine the distance-to-target pattern of detections while accounting for retinal eccentricity and peripheral discriminability. Step 3c predicts reports concentrated near the landing X, with spillover declining with distance. Full-field Step 5c predicts detections beyond that region, although the required change magnitude may increase with eccentricity. A near-target excess superimposed on a peripheral tail adjusted for discriminability would support contributions from both mechanisms.

### Why Step 4 Cannot Be the Source of the Bottom-Up Attention Request

During Step 4, the Modeler uses multisensory input to *update* both the Concrete World Model and the Focal Target Information stream. Focal-target processing integrates relevant inputs for each selected target and supplies the resulting Focal Target Information to the Controller$_{whole}$. Broader sensory-field processing *updates* the Concrete World Model's current sensory contents, including non-focal regions available to the Modeler$_{schema}$ for generating diffuse-awareness qualia.

Step 4 alone lacks the required comparison state. By saccade end, the Concrete World Model contains the post-saccadic sensory state, but Step 4 has not buffered the pre-saccadic diffuse visual content. Therefore, Step 4 cannot generate a bottom-up attention request from a non-focal permanent change whose transition occurs entirely during the saccade.

By contrast, Step 5b holds the Quale World Model's current-sensory visual portion at pre-saccadic values under either Step 5 architecture. Only full-field Step 5c then compares each held non-focal quale with a candidate post-saccadic quale generated from Concrete World Model content updated in Step 4 before replacing the held quale with the candidate.

Thus, once Steps 1, 3c, and 4 are excluded, a bottom-up attention request triggered by a non-focal permanent change occurring entirely during the saccade would, within the two proposed architectures, require full-field Step 5c.

### Evidence for Distinct Representations in Cross-Saccadic Consistency Checking

Step 4 is also excluded on representational grounds: it would have to preserve pre-saccadic Concrete World Model visual content while constructing the post-saccadic state and then compare the two. But a Step 4 mechanism performing that discrepancy-detection comparison to tune the Modeler would effectively be operating as the Modeler$_{schema}$. That would collapse the distinction between visual Concrete World Model content and diffuse-awareness visual qualia, making the Concrete quale-conversion process for that content an identity mapping.

That broader identity-mapping hypothesis conflicts with the evidence discussed in "Distinct Qualia from Distinct World Model Components." If visual concrete-format World Model content were generally identical to visual qualia, then the content preserved in Memories of the Concrete World Model would make recalled visual targets perception-like by default. They are not: recalled visual targets often lack perceptual vividness and vary widely, from aphantasia to hyperphantasia. Therefore, the evidence weighs strongly against treating visual World Model content and visual qualia as generally identical and supports a distinct Quale World Model.

### Novelty and Scope of the Experiment

To our knowledge, no prior study has used attention capture to infer a bottom-up attention request from a non-focal cross-saccadic consistency check. Prior paradigms changed saccade targets and relied on explicit reports. Here, changes to objects lacking pre-saccadic focal representations can distinguish full-field Step 5 from focal-target processing in Step 3 or focal-target Step 5.

A sufficiently sensitive null result would reject full-field Step 5c while leaving focal-target Step 5 and the broader theory viable but largely untested by this experiment. The experiment therefore tests whether qualia-based cross-saccadic checking extends beyond focal targets, not whether focal-target checking occurs. Under either architecture, qualia remain the $\text{Modeler}_{\text{schema}}$'s consistency-checking representations, returning us to the Hard Problem.

## A Proposed Solution to the Hard Problem of Consciousness

Suppose science has precisely determined all the functions the brain performs to produce behavior. Chalmers (1995) defines the Hard Problem as asking why "the performance of these functions" is accompanied by experience. In this theory, such a functional account could include the main behavior-producing pathway from sensory input through the Modeler and World Model to Focal Target Information and the Controller. We agree that even a complete description of that pathway would not, by itself, explain why any state is experienced. The proposed solution instead concerns the physical identity of the experiencer, experiencing process, and experienced content—a unity not captured by that pathway.

### Experiential Unity: Subject, Process, and Content

The Hard Problem may arise partly from assuming an unnecessary additional relation between qualia and an experiencer. In this theory, qualia do not produce feelings in a separate internal observer. Instead, conscious experience is physically identical to the unified subject, processes, and content of one system, the $\text{Modeler}_{\text{schema}}$, which plays all three roles:

- **Experiencer**: The $\text{Modeler}_{\text{schema}}$ is the minimal phenomenal subject—the only agent in which qualia occur. Because it is functionally incorporated within the $\text{Modeler}_{\text{whole}}$ and $\text{Human}_{\text{whole}}$, the $\text{Human}_{\text{whole}}$ can attribute the $\text{Modeler}_{\text{schema}}$'s experiences to itself (Figure 3), explaining why we identify ourselves as conscious.
- **Experiencing process**: The three quale-conversion processes, together with their updating of the Quale World Model and consistency checks on it, constitute experiencing—they are not merely associated with experiencing.
- **Experienced content**: The Quale World Model constitutes the content of experience, constructed from current Concrete World Model contents for diffuse awareness and from the Focal Target Information stream for focal experience. Each quale is a physical internal representation, not a non-physical substance.

None of these three components are part of the main behavior-producing pathway. Their unity requires no further feeling or observation of qualia. In other words, a person's conscious experiencing is physically identical to the activity of that person's $Modeler_{schema}$ as it generates and updates qualia within the Quale World Model and checks their consistency, whether during wakefulness or dreaming. Consciousness is absent when the $Modeler_{schema}$ is not performing these experiencing processes.

By contrast, Graziano's (2016) Attention Schema Theory places the attention schema within behavior-guiding machinery: it represents a physical self and object connected by attention and explains why awareness seems non-physical: the attention schema represents this relationship without representing its underlying neural mechanisms. The $Modeler_{schema}$ Theory instead identifies the $Modeler_{schema}$ itself—not a representation—as the physical minimal phenomenal subject; its experiencing processes and Quale World Model likewise lie outside the World Model's represented universe. None of these appear as represented objects or processes within the World Model's universe.

## The $Modeler_{schema}$ as a Self-Contained Universe

World Model content can become focal and pass through Focal Target Information to the $Controller_{whole}$ for action and report. Quale World Model content remains private: qualia never leave the $Modeler_{schema}$ as content, though the processes generating and using them can affect the system through Modeler refinement, Appraisal Emotion signals, and, under full-field Step 5c, bottom-up attention requests. Compared with ordinary sensory processing, $Modeler_{schema}$-generated Appraisal Emotion signals have subtler behavioral effects, chiefly enabling reports about characteristics of qualia inaccessible to the $Controller_{whole}$. Full-field Step 5c requests could observably redirect focal attention in the proposed experiment but should otherwise be rare in normal operation. A sufficiently sensitive null result would eliminate only the full-field Step 5c bottom-up-attention route within the proposed architectures; Appraisal Emotion-based reports would remain possible, and focal-target Step 5 viable but unconfirmed.

Yet the $Modeler_{schema}$ remains a black box to the rest of the system. Phenomenology exists only within this self-contained universe and never appears as a World Model object or property, helping explain its apparent absence from ordinary physical descriptions of the world.

The $Controller_{whole}$ constructs first-person reports from Focal Target Information and associated $Modeler_{schema}$-generated Appraisal Emotion signals, not from qualia themselves. It receives report-enabling information about experience but no content-preserving access to qualia. Thus, the $Controller_{whole}$ can report "I experience the recalled face as vivid" because it receives both recalled-face target information and a "*vivid*" Appraisal Emotion signal, while the corresponding experience occurs only within the $Modeler_{schema}$. In aphantasia, task-usable recalled visual information may remain available as a focal target even when its visual quale is absent; an "*absent*" Appraisal Emotion signal can then support a report of no visual experience despite the $Controller_{whole}$'s access to that target information.

Visual diffuse awareness most clearly shows that the $Controller_{whole}$ is not the conscious subject. It can report only focal-target content, yet in the diffuse-awareness exercise the concept of diffuse awareness becomes an abstract focal target.[21] The Modeler$_{schema}$ continuously converts the concrete visual field into diffuse-awareness qualia and can attach the Appraisal Emotion "*present*" to that abstract target when it is *understood* as referring to the diffuse-awareness quale field, allowing the Controller$_{whole}$ to report "I experience diffuse awareness" through the I/Me/My$_{model}$ even though it does not experience the field itself. Indeed, this paper is a Controller$_{whole}$ construction using the terms "experience," "consciousness," and "qualia" without direct access to what they refer to.

As part of the experiencing process, qualia-based consistency checks may calibrate the Modeler during development and recalibrate it after later sensory changes or impairments, such as an acquired scotoma. In an adult, if a stroke selectively disabled the Modeler$_{schema}$'s quale-generating and accuracy-monitoring functions, ordinary behavior might remain largely normal even without experience. Targeted tests, however, should reveal missing consistency checks and Modeler$_{schema}$-generated Appraisal Emotion signals such as "*vivid*," "*faint*," "*absent*," or some uses of "*surprised*"; yet Modeler-generated Appraisal Emotion signals would remain. Under either Step 5 architecture, such a person would be a functional approximation to a philosophical zombie. If full-field Step 5 is experimentally supported, missing full-field detections would provide an additional sign of the impairment. Disabling the Modeler$_{schema}$ in infancy, however, would substantially impair Modeler monitoring and development, including the calibration of perceptual processing. Avoiding that outcome is the proposed evolutionary function of the qualia-based monitoring process and, because that process constitutes experiencing, of conscious experiencing itself.

## Why We Misattribute Consciousness: The Control Hierarchy Illusion

The Human$_{whole}$ can claim consciousness because it functionally incorporates the Modeler$_{schema}$. In everyday life, however, the Controller$_{whole}$ identifies itself with the Human. Its window on the currently sensed world consists of the portions of the Focal Target Information stream representing Concrete World Model focal targets, which it treats as the world itself despite their representing only a focal-target subset of the sensory field available to diffuse awareness. Because the Controller$_{whole}$ selects actions, sets Modeler goal priorities, requests attention, produces language, and controls the body—essentially everything a Human chooses to do—it naturally identifies itself with the Human. Because it does not understand the distinction between the Human and Human$_{whole}$, it also takes that Human to be conscious.

The Controller$_{whole}$ thus mistakes World Model content received through Focal Target Information for direct conscious experience of the external world. The diffuse-awareness exercise exposes this mistake because the broader visual field remains consciously present in the Modeler$_{schema}$, and therefore attributable to the Human$_{whole}$, but outside the Controller$_{whole}$'s direct access even though it identifies with the Human.

This division explains why consciousness seems alien to the physical world. The Controller$_{whole}$ is the embodied, acting, speaking self represented by the Body$_{model}$ and I/Me/My$_{model}$; the Modeler$_{schema}$ is the unrepresented minimal phenomenal subject. The $\text{Modeler}_{\text{whole}}$ constructs the World Model, while the $\text{Modeler}_{\text{schema}}$ contains representations derived from World Model content without being represented there. This one-way relationship makes consciousness seem external to the physical world despite its implementation in the brain.

[21] A reader may have experienced diffuse awareness throughout life without previously having the concept of diffuse awareness available as an abstract focal target.

To our knowledge, existing theories of consciousness do not identify a comparably precise, body-independent functional control agent as the minimal phenomenal subject. Many theories emphasize focal-access mechanisms[22] and therefore explain focal experience more readily than diffuse awareness. Others may accommodate diffuse awareness but do not clearly distinguish task-usable recalled visual information from its visual quale, as required to explain aphantasia.

The $\text{Modeler}_{\text{schema}}$ does not produce experience for something else: the unity of its subject, process, and quale content is physically identical to conscious experience.

## Conclusion

The $\text{Modeler}_{\text{schema}}$ Theory identifies a specific schema agent as the minimal phenomenal subject and the only agent in which conscious experience occurs. By physically identifying conscious experience with the unity of experiencer, experiencing process, and experienced content, it offers a path toward closing the explanatory gap: no separate observer or additional feeling relation is required. Qualia are functional internal representations used to calibrate and optimize the Modeler.

If confirmed, the experiment would support an ongoing role for diffuse-awareness qualia in perceptual stability and calibration. Under adequately sensitive and controlled conditions, failure to detect permanent peripheral changes, greater detection of matched-magnitude temporary changes, or detections confined to the saccade landing region would reject full-field Step 5 while leaving focal-target Step 5 and the broader theory viable but largely untested by this experiment.

## Acknowledgment

We thank David McFadzean, whose sustained feedback on many drafts improved the theory's structure, terminology, and cybernetic foundations.

## Declaration of AI-assisted tools

The author used AI-assisted tools for wording and literature searches; independently verified cited sources; edited AI output; and takes full responsibility for the content.

[22] Examples of theories emphasizing focal-access mechanisms include Global Workspace Theory, Global Neuronal Workspace, Higher-Order Thought, and Attention Schema Theory.